\documentclass{article}

\usepackage{arxiv}

\usepackage[utf8]{inputenc} 
\usepackage[T1]{fontenc}    
\usepackage{hyperref}       
\usepackage{url}            
\usepackage{booktabs}       
\usepackage{amsfonts}       
\usepackage{nicefrac}       
\usepackage{microtype}      
\usepackage{lipsum}
\usepackage{graphicx}
\graphicspath{ {./images/} }

\title{ Communication Aid for Non-English Speaking Newcomers }

\author{
 Munira Al-Ageili \\
 Department of Computer Science \\
  University of Regina\\
  Regina, SK S4S 0A2 \\
  \texttt{munira.al-ageili@uregina.ca} \\
   \And
 Malek Mouhoub \\
  Department of Computer Science \\
  University of Regina\\
  Regina, SK S4S 0A2 \\ 
  \texttt{mouhoubm@uregina.ca} \\
}

\begin{document}
\maketitle
\begin{abstract}
This research work is intended to assess the usability of Pictogram symbols and other visual symbols in an audio-visual strategy to facilitate and enhance the use and learning of English as an additional language for Arabic-speaking Syrian refugees, with a potential for generalizing the process to speakers from other linguistic backgrounds. The adopted software for the project is PICTOPAGES – a versatile tool with 2,200 symbols, 78 animated symbols, and the potential for customization with photographs, thus augmenting its capability for personalization and relevance.  While PICTOPAGES is the intended basis for this research, the concept and software will be adapted and modified as may be required.

PICTOPAGES includes text, recorded speech, and symbols and is currently available for iPad. In the future, it may be adapted for use on iPhone. A preliminary design using PICTOPAGES has been created for this research.  The focus group includes, but is not limited to, newcomers who may have limited to no English skills, limited resources, limited education, and potentially limited literacy in their native language, and perhaps high levels of distraction / frustration related to their recent experiences.  Enhanced communication capability and confidence should enhance the participants’ employment potential. Extensive interaction with respect to communication requirements, selection or development of readily understandable symbols, and real-world testing would be undertaken with an intended user group. A potential subset of the focus group could involve members of the refugee community that, in addition to English language limitations, also have developmental or acquired disabilities that affect their ability to communicate verbally (per the original intent of the software).
\end{abstract}


\section{Introduction}
The use of pictograms to provide communication assistance to individuals with developmental or acquired disabilities which impair their verbal communication capabilities is a relatively common and well-known strategy.  Low-tech to high-tech aids are available, from symbol-based communication boards (e.g., printable from software applications) to symbol- and picture-based applications on mobile devices and specialized, dedicated messaging devices.  Similarly, there has also been some use of pictograms to assist in the teaching of alternate languages \cite{Takil2016}, interest regarding their use in discordant language situations for specific needs (e.g., delivery of healthcare services) \cite{clawson2012using}, and some use to assist refugees who do not speak the language of their host countries \cite{Miller2016,McNeill2015}.  A wide variety of symbols and pictures have been employed under the very generic label ''pictogram''.

The proposed work attempts to use the pictogram approach, most commonly used for special needs individuals, in the development of a practical, mobile communication aid on an iOS device that would be useful to the Arabic-speaking refugee community in Saskatchewan.  Communication needs and feedback on symbol selection would be provided by target group participants in the project.  Because the audio prompted via symbol selection would be provided in both English and Arabic, users would have immediate confirmation that their symbol selection / associated ''translation'' into English of what they are trying to say is correct.

The novelty of the proposed work is with respect to the type of communication aid to be developed – i.e., pictogram-based, loaded on an iOS mobile device, with bilingual audio output – for what largely remains a non-traditional user group for pictogram-based communication systems, built on the basis of user input for both content (vocabulary:  the communication needs) and symbol selection, and validated with the users in ''real-world'' situations.  It should also be noted that the intended platform, on which the proposed application would be built, allows for subsequent addition of custom text, recorded audio, and imported images by, or on behalf of, the user.

In 1976, Maharaj began working on the development of the Pictogram Program, a visual strategy for communication for nonverbal individuals.  Pictogram Symbols, a symbolic language system, has been in use since 1980.  It provides a communication strategy for children and adults with disabilities – for example, cerebral palsy, autism, stroke, Alzheimer’s, Parkinson’s – that have impaired their verbal communication capabilities widely \cite{Pictocom2017}.

Pictograms, consisting of a white symbol – a simplified “picture” – on a black background, can illustrate objects, concepts or actions and provide adaptive opportunities for communication for those who require such assistance for communicating with those around them, as well as providing a platform for the creation of classroom materials to benefit those individuals. Pictocom International has Pictogram Centers in Canada, as well as in Sweden and Japan, where Pictograms are used widely \cite{Pictocom2017}. The program has gained international acceptance over the years and in its basic and applied forms continues to be a significant strategy for nonverbal individuals in many parts of the world.  SIT, Sweden and J-PIC, Japan, are major partners in the development of Pictogram materials.

Communication software applications have been developed based on the use of pictogram symbols. They include {\bf PICTOCOM Print and Speak (PICTOCOM PS)}, using pictograms developed by   Maharaj. This tool was developed and is available for desktop Mac and Windows computers. The intent of {\bf PICTOCOM PS Print} is to create customized communication boards for individuals who are not able to communicate verbally and also unable to use a communication device.  The application can be used to create lists (e.g., for groceries), restaurant orders, and activity schedules.  Although communication boards can be created for all ages, a major focus is for the creation of education material for non-verbal children, including learning the alphabet, spelling, and vocabulary development.

{\bf PICTOCOM PS} Speak allows for voice attachment to each symbol; optionally, a customized voice recording can be used instead of the system voice.  Pre-recorded lead-in statements (e.g., “{\it I would like $\ldots$ }“) can be used or these can also be customized.  The application can function in two ways.  In its simplest form, touching a symbol displays the text and then touching “Talk” produces the spoken response.  Alternatively, a lead-in can be selected and then the desired pictogram (e.g., “{\it I would like $\ldots$  an apple}”); pressing “Talk” then produces the spoken lead-in plus pictogram.

In 2014, the University of Regina established the George Reed Centre for Accessible Visual Communication (GRC), funded by the George Reed Foundation, to support Pictogram research and development and to assist exploration of other means of visual communication to help people with physical or intellectual disabilities.  The GRC funded a project for the {\it Investigation of the application of Pictogram Symbols for children and adults who require a visual strategy for communication and the creation of applicable software programs}.  This project was a collaborative effort between Dr. Maguire, Professor of Computer Science at the University of Regina, Subhas Maharaj, originator of the Pictogram Symbol system and Senior Speech-Language Pathologist, Community Health Services, Heartland Health Region, Rosetown, SK, and Postdoctoral Fellow Dr. Munira Al-Ageili, University of Regina.  An Augmentative and Alternative Communication (AAC) application was created for iOS mobile devices, such as iPhone and iPad, rather than for dedicated speech generating devices (SGD).  The software for the project is {\bf PICTOPAGES} – a versatile software with 2,200 symbols, 78 animated symbols, and the potential for customization through photographs, thus augmenting the personalization and relevance of the learning process.  The app enables users (or their caregivers) to create, share and interact with Pictograms.  The app allows customization of the page layout, the addition of pictograms with text and user recorded audio, and launching of the PictoPage Viewer to communicate from the iPad via touch.  The key features are:
\begin{itemize}
\item	Create, customize, and manage PictoPages.
\item	Access the online Pictogram Library and import images from a photo library or camera.
\item	Include custom text and record and attach audio to pictograms and imported images.
\item	Launch and interact with PictoPage on an iPad.
\item	Print and share PictoPages via email.
\item	Create and use PictoPages with or without an internet connection.
\item	Sync locally stored pictograms with the online Pictogram Library.
\item	Login and access one’s PictoPages from other iPads.
\end{itemize}

 Maharaj launched PICTOPAGES at the November 2016 American Speech-Language-Hearing Association (ASHA) Convention in Philadelphia; 1,300 pieces of information about the app were distributed during the 2½-day convention and PICTOPAGES was referenced by some of the later presenters.

Although various visual aids may be commonly used as part of language instruction, including foreign language instruction, the use of Pictograms as a communication tool for learning language appears to be somewhat limited. Maharaj has suggested that ``{\it there has never been a specific need}'' to use Pictograms with individuals who do not have impaired verbal communication capabilities; ``{\it however, classes of students have been included in activities and projects to apply the concept of inclusion – this was observed in Sweden and Japan}''.  Falck \cite{Falck2001} states ``{\it Pictogram is used at many pre-schools today where mentally retarded children have been integrated into classes. In these cases, Pictogram has become a common model for all the children when it comes to learning to read and write}''.  Hart \cite{hartancient} planned to teach a class of Spanish immersion kindergarten students in the U.S. with the assistance of pictograms and mixtures of words and pictures. However, at least part of the decision to use pictograms appeared to be for inclusion purposes based on the presence of developmentally delayed student in the class (per the concept of inclusion, as mentioned by Maharaj).

Takil \cite{Takil2016} discusses the work done to use the pictogram method for teaching Turkish as a foreign language.  It was stated that while some pictogram-based materials for the teaching of foreign languages can be found in the literature, such materials are scarce. The approach described used a bold, high impact, text-based style of pictograms, where words were shown in a variety of fonts, sizes, alignments, and colors with added images, such as a leaf growing out of a letter, a cartoon drawing of a sad man sitting on a letter (with pools of tears below), a backdrop of a sun peeking above the clouds, letters with eyes and mouths, and ''cold'' blue letters (one with a toque) in what appears to be a snowfall.  These teaching aids were intended for use in a classroom setting.  It was  reported that this approach was not only effective for teaching a Turkish vocabulary, but increased both the ease of learning and the retention of that vocabulary.

This use of pictograms in the form of (literally) word drawings is a very different approach to pictogram use than is being contemplated in the work being proposed here, which will use a more conventional, symbolic pictogram style, with limited or no text, and augmented with spoken audio.

Clawson \cite{clawson2012using} discusses the use of pictograms with respect to providing health care in the context of US Navy exercises intended to provide training for humanitarian and disaster relief to U.S. military, NGO, and other associated personnel.  Communication between those providing and those receiving medical care was problematic, due to either a lack of skilled translators or to translators who had little or no knowledge of medical terminology and practice.  In most cases, it was not possible for the English-speaking medical personnel to determine whether what was being said was being correctly translated.  In some cases where NGO translators were available to monitor local translators, it was found that information was not being accurately provided, with some of the information being highly inaccurate.
Given this situation, there was an identified need to provide medical staff with alternate methods of communication; additionally, those methods would require testing for effectiveness and validity before being put into practice. One potential method identified was the use pictograms representing common medical conditions / symptoms.  To determine whether such pictograms would be capable of meeting the 85\% level of accuracy specified by the American National Standards Institute (ANSI), thirty-six images (including three duplicates) were provided to medical personnel for interpretation.  It was found that over 75\% of the (unique) images met the ANSI criterion.  This suggested that the use of pictograms could be a viable communication method when medical staff and patients do not speak the same language.

Clawson \cite{clawson2012using}   describes a 2006 study by Houts and others that looked at the effects of pictures for health communication.  The study found that people’s ability to understand medical information was often hampered because of illness, stress, emotional distress, and unfamiliar terminology.  Except for illness, the other factors claimed to negatively affect the processing of information are also relevant to the Arabic-speaking refugee population which is the target group of the project.  The study authors suspected that a combination of pictures with text would improve medical conversations.

Clawson\cite{clawson2012using}  also describes the work of Delp and Jones, who suspected that pictures (“cartoon illustrations”), in combination with verbal and written instructions, might improve medical staff-patient communication, and therefore improve patient understanding of discharge instructions.  The authors reported that patients provided with the picture-enhanced instructions found them easier to read than was the case with patients provided with only the written instructions.

With respect to the cross-cultural usability of symbols, it is pointed out   that a South African population did not understand the U.S. pictograms used on prescription medications \cite{clawson2012using}.  Clawson  \cite{clawson2012using} states that “Dowse and Ehlers found that graphic material, once felt to be a universal language, has distinct cultural biases” (p. 294).  These authors recommended that the pictograms to be used should be designed in conjunction with the target population and that the pictograms should be tested to generate feedback for modification in order to meet the specified level of identification accuracy.


With respect to Clawson’s  work \cite{clawson2012using} regarding the validation of twenty-six of the thirty-three unique images, it is stated that the results should not be interpreted to apply generally to other cultures / languages; the same process of interpretation validation would be required for each different language or cultural group.  It is noted for this research that the symbol library that would be created may serve as a baseline for comparison of interpretation with newcomers speaking other languages and from other cultural groups.

With reference to refugees from Syria and elsewhere arriving in European cites, it is described how Bauer, from a Vienna-based design studio \cite{Bauer},  approached the NGO that ran the Wien Mitte refugee camp \cite{Miller2016}. The camp’s volunteer translators had gone and ``{\it the remaining visual communication system amounted to some make-shift signs taped to the walls.  It was chaotic $\dots$ Without a common language between all of the refugees, there was no comprehensive understanding of what the written signs were saying.  The easiest solution $\ldots$ was to design a set of icons that could be easily understood across cultural lines:  a pictogram language}''  \cite{Miller2016}. Icon design started with the basic necessities – health services, bathrooms, and food – with the designers asking for feedback from the translators and from the refugees, themselves. Based on that feedback, modified versions of the icons were created.  The resulting system was the result of collaboration between these three groups.  This collaborative approach is similar to what is being proposed here.

Miller claims that ``{\it Bauer’s pictogram system   differentiates itself with an emphasis on respect and symbology that reflects both the cultures of those seeking asylum and those acting as host}'' \cite{Miller2016}.  Examples provided to illustrate this approach include symbol placement which takes advantage of the left-to-right reading of European languages vs the right-to-left writing of the Arabic-speakers, where both parts of a paired symbol can be seen as coming ``first'', and the icon developed for ``woman'', which may be viewed as either a woman with particularly long hair or as a woman wearing a headscarf.

Although Bauer’s system is available \cite{Bauer}, it continues to evolve via collaboration with others, such as a typographer who digitized the icons and made them available for use in Berlin refugee camps, and an organization which was creating an app for refugees to provide guidance on issues such as schools and affordable food and housing.

In response to an on-line question regarding the use of symbols in schools for students with English as an additional language (EAL), McNeill  \cite{McNeill2015} responded that some Scottish schools were quite advanced with respect to the use of symbols to support communication and learning for all students.  She acknowledged, however, that other schools may have a more limited awareness of this approach, with any use directed only towards the support of individual students.

The approach described by McNeill \cite{McNeill2015}  is relatively low-tech:  essentially, use of software for printing symbol-based communication boards and the selection and creation of associated teaching resources.  She advocates the development of personalized vocabularies, rather than relying on ``off the shelf'' communication boards.  The extent of personalization advocated by McNeill isn’t clear; in the context of a teaching environment, she may indeed be suggesting customization down to the individual level.  The work being proposed here would result in a vocabulary and symbol set customized for a particular group – rather than specific individuals – to develop a common communication application.  However, because the vocabulary and pictogram set developed would essentially be the summation of individual participant inputs / feedback, communication needs and symbol understandability at the individual level should largely be addressed (within the participant group) by the application developed.

McNeill  \cite{McNeill2015} also identified a company that has produced a symbol-based communication board for refugees with little or no English.  These boards have Arabic and English text in addition to the symbols, which limits their use with refugee children whose schooling has been interrupted.  In such cases, understanding the messages on the board becomes completely dependent on accurate interpretation of the symbols, themselves; however, this is probably made more difficult by the fact that the symbols were presumably not designed for standalone use.

A guide published by Special Education Technology – British Columbia \cite{SETBC2008} claims that it is a myth that AAC is only for those who cannot communicate verbally.  The guide’s position is that AAC can be used by a variety of people.  Having said this, however, the guide primarily references individuals with very significant communication barriers such as cerebral palsy, autism, cognitive deficit, and multiple disabilities and discusses selection of messages (access) with respect to factors such as an individual’s motor skills, physical limitations, vision, and hearing.  This focus is further reinforced by the recommended involvement of an occupational therapist to provide access guidance.  These are users whose special needs are far more severe and complex than simply not understanding the English language.
The guide describes an AAC system as comprised of three parts:  a means of representing things or concepts (i.e., the symbols or pictograms), a means to select or access those symbols, and a means to transmit the information.

In cases where the user is required to point to or touch the message – or the symbol representing the message, this is described as ``direct selection''.  Per the guide, this is the preferred access method because it is simple and quick.  The communication aid to be developed through the work proposed here uses direct selection of the symbols by the user.
 
The guide goes on to provide examples of symbol sets and both single and multiple message devices, but again, the focus is on serving a very different (and more traditional) user group than the proposed project.

In a study on immigration and information and communication technology (ICT), co-funded by the European Commission, a project to address of an apparently increasing number of EU citizens with limited literacy skills is described \cite{Kluzer2013}.  It is noted that the success rate of adult illiterates attending literacy courses was low.  A study by Alpha-Beta is described which examined the use of mobile phones loaded with nine modules designed to provide adult education services to this target group.  The modules included material on healthy living, learning of words and numbers, prevention of sexually transmitted diseases, and family-school relations.

The education modules were small software packages, installed on mobile phones, intended for independent learning by individuals at times of their choosing.  The different modules were found to have different levels of success:  a ``picture story'' about disease prevention was popular but one intended to provide shopping assistance, which required the pressing of specific phone buttons, was found to be too complicated for effective use.

Again, while there are elements in the above description that are shared with the proposed work – immigrants, software loaded on mobile phones, pictograms – the Alpha-Beta study described was very different:  in focus (adult education vs an aide for communication and English learning enhancement), in target audience (illiterate immigrants vs simply non-English speaking ones), and in the type of software provided on the mobile phones.
 
Section \ref{sec:objective} lists the objectives   for this research work. Section \ref{sec:methodology} describes the methodology that we propose to achieve these objectives. Finally, concluding remarks are highlighted in Section \ref{conclusion}. 
\section{Objectives}
\label{sec:objective}
 
 The main of objective of this work is to develop a pictogram-based communication aid for Arabic-speaking newcomers.  The communication 	aid will allow selection by the user, based on pictogram recognition, of the message to be 	communicated.  Upon selection, the tool will provide recorded audio output of the corresponding word, 	phrase, or sentence in both Arabic and English.

The above objective can be structured in the following sub-objectives.
\begin{enumerate}
    \item Vocabulary
\begin{enumerate}
    \item Build a vocabulary that consists of communication needs of Arabic-speaking newcomers that is
           based on needs as identified by that group rather than on assumptions about their needs.
   \item  Generalized vocabulary:  by design, the vocabulary will cover a range of subjects useful for a
     variety of users rather than being intended for the use of a specific group with specific needs
     (e.g., children with autism), as is commonly the case, or for a specific subject area (e.g., to assist
     in the provision of medical care).
     \item  Reusability of the vocabulary in this or some other communication aid for other non-English
     speaking newcomers – i.e., people in similar circumstances with similar needs, regardless of
     their native language.
\end{enumerate} 

\item Symbols
\begin{enumerate}
    \item  The selected symbols will be validated with the user group regarding interpretation and replaced
	            or modified, where necessary.  Symbols which cannot be readily understood by users have little  
	            value; however, it is unrealistic to assume that all users will understand all symbols.  An
	            acceptable level of symbol understanding across the user group will be determined.
\item    It is believed that there is a cultural component to symbol appropriateness and understanding. Therefore, this validated symbol 
set may provide a baseline for comparison with newcomers speaking other languages and from other cultural groups.
\end{enumerate}

\item   Typically, pictograms have been used as a teaching aid for an instructor in certain classroom
        settings.  However, this tool is intended for independent use.  Using the tool in real-world  
        situations, hearing the messages in both English and Arabic, and seeing immediately whether the     
        intended message was understood by the receiver, is anticipated to enhance users’ acquisition of
        English.  Further, because smart devices are typically in frequent use for entertainment and  
        ``look-up'' purposes in addition to communication, the availability of the tool on a user’s smart  
        device may encourage independent exploration of the language content between occasions of real- 
        world use.

\item    Depending on the individual members of the intended target group who choose to participate in the  
       project, it may be possible to test the communication aid in both the ``additional language'' and 
       traditional special needs / verbal impairment realms, if some participants fall into both categories or  
       subsequently recruit such individuals known to them.  Individuals who face both challenges may 
       find this approach of particular value, although optimum use by this subgroup might require some 
      degree of customization.

\item    The working communication aid, in addition to providing actual and on-going value to the target 
       community – Arabic-speaking newcomers – past the life of the project (particularly if there is
       subsequent interest in continuing to build on that platform), should be able to serve as a template for  
       subsequent use with other languages:  the vocabulary (based on identified needs), many of the  
       pictograms, the English audio and the actual software should be re-usable; for the most part, other 
       than possible changes to the pictogram set, what would need to be created would be the   
       corresponding audio in the new language.

\item   Follow-up comparison of this approach with more complex / powerful approaches towards  
       Augmentative and Alternative Communication (AAC) – such as linked ontologies making use of 
       formal semantics – may provide interesting insights, especially from the perspective of the amount  
       of development effort vs actual utility to users.

\end{enumerate}

\section{Proposed Methodology}
\label{sec:methodology}
Developing a communication aid for individuals in daily need of such assistance is very much a user-centred design process.  A great deal of the overall project effort – and a critical element in its success – will be the collection, compilation, and analysis of detailed information from participants drawn from the intended user group.  Input from the group would be used to specifically identify the situations / interactions where verbal communication assistance is most required, what useful assistance would look like, and to precisely identify the information that they need to be able to convey in each case.  A clear understanding of the mobile device capability / familiarity of the participants, constraints regarding the cost / sophistication of devices available to them, and how use of the communication aid would successfully fit into their daily interactions would be developed.  Being able to recruit a cross-section of users would facilitate definition of a suitably broad set of functional requirements to guide development of an end-product that would have value across the entire target community.  Further, the process is multi-staged: 
\begin{enumerate}
    \item  it must involve frequent consultation with participants, and 
    \item on-going testing of the design throughout the process, 
    \item finally  culminating in real world validation, both researcher-observed / mediated and independent.
\end{enumerate}

To date, one Syrian family has been participating in early work on the project (in order to help develop scope and process); there have been several meetings with the family.  Following appropriate introductions and familiarization, the intent of the project, the communication challenges they face, the information-collection forms, and the PictoPages app were discussed with the family.  A small demo (custom symbols, with English and Arabic audio) that had been constructed was shown to the family.  Some areas of interest and associated important phrases / sentences were identified with the family and documented; a few symbols were selected in discussion with the family and the associated phrases, in both Arabic and English, recorded with the assistance of one of the daughters.

Our general methodology can be achieved through the following tasks.

\subsection{Recruitment – Identification of Potential Participants}
Information about the existence and intent of the project first needs to be disseminated among the local refugee / immigrant community.  The current thought is that a combination of 15 – 20 individuals and family groups (each consisting of a number of individuals, some of whom may be children) may be the appropriate number of participants.  The intern has personal connections to the local Middle Eastern / North African community, thus personal networking can serve as part of this process.  However, assistance would also be solicited from groups and organizations serving this community – in particular the Regina Public Library, the Regina Region Local Immigration Partnership, and the United Way.  Formal understandings or agreements – if any – required by these groups regarding consent and process may be needed.  Particularly for prospective participants identified via personal connections, the researcher may initiate a brief, personal contact (e.g., by phone).  In other cases, however, initial contact may need to be mediated via the assisting organization or group. Other local organizations may be approached for assistance on an ad hoc basis.

\subsection{Recruitment – Securing Participation}
Having initiated contact with potential participants, the researcher would conduct an introduction / initial interview, preferably in person (or, where necessary, by phone), not by email.  The intern, who speaks fluent Arabic, would explain the intended process and try to get a feel for the parties’ needs and what they wish to achieve.  Potential participants would have to be satisfied that there is value to them in participating; in turn, the researcher must be satisfied that the potential participants can contribute – i.e., that they have the need, time, and capability to participate (including the likelihood of their remaining in the community over the course of the project).  Importantly, participants and researcher must both feel comfortable / safe in their interactions.  In order to make the application as broadly useful as possible, participants with a range of English skills, education, family situations, mobile device familiarity, and specific needs would be sought.  Participants would need to understand and agree to their expected contributions / roles over the course of the project, the anticipated time required, and the project’s overall duration; in turn, the researcher needs to be able provide the flexibility to accommodate commitments and events affecting the participants’ lives, as these evolve.  Any consent from the participants per University or Mitacs requirements would need to be obtained.

\subsection{	Interviews:  Needs / Tasks – Information Collection}

		Develop, refine, and structure a comprehensive listing of subject areas and the specific 			communication requirements (words, phrases, sentences) in those areas.
			\begin{verbatim}
		e.g., banking 
-	I would like to open a savings account.
-	I would like to cash a cheque.
-	I would like to pay my credit card.
		e.g., medical
-      I have a pain in my ….
	-      I am not sleeping well.
	-      My son/daughter has a fever.
		e.g., getting around
	-     Can you please tell me where I am?
	-     Can you please tell me how to get to …
	-     Can you please tell me where to catch the bus?
	\end{verbatim}

As communications requirements, including both subject areas of concern and specific words, phrases, statements, questions, $\ldots$ etc, are identified by the participants, via a combination of written forms and live discussion (the combination of which may need to be adjusted to suit individual participants), this data would be categorized, sorted, and refined.  As the language information accumulates, the input from the various participants would be integrated; the researcher would begin organization / structuring of the collected information with the goal of facilitating both the building of the content and the ability of users to intuitively and rapidly navigate that content.  The existing paper forms whose purpose is to assist in the collection of such information would be modified as needed to increase their usability and value in this role; adaptation to mobile devices may need to be considered.

\subsection{	Symbol Identification / Development / Participant Review}

The researcher would work closely with the participants to identify pictograms or other symbols that could be matched to the identified words, statements, etc.; such symbols must be readily and widely understood by members of the intended user group; where such symbols are not available, the researcher may need to develop suitable symbols using photographs, custom graphics, and (if necessary) limited text in Arabic  Tentative symbol selections must be broadly reviewed.  To be of value, the symbols must be as obvious and unambiguous as possible and the content lists that use these symbols should be rapidly ''scannable'' by users; i.e., the visual symbols must enhance the user’s ability to quickly find and access content as they are engaged in communication.  In some cases, symbol selection must recognize that some symbols that may be readily understandable to North American or European populations may not be obvious – or in some cases, appropriate – to individuals from other cultural backgrounds.

\subsection{Building of Content (using PictoPages) / On-going Review / Refinement}
The elements to be communicated (words, statements, questions, etc., in both Arabic and English), along with the pictogram or other symbol to be associated with each language pair, would be reviewed and organized to facilitate construction of the content (active pages) using PictoPages.  Symbols would be selected from the PictoPages library, imported from publicly available sources, or imported from the collection of symbols created by the researcher or collaborators, as required.  Audio content, in Arabic and English, would be recorded / linked to the associated symbols.  Symbol resolution and clarity would be reviewed and deficiencies addressed.  Recorded audio content (in both languages) would be reviewed for correctness (accuracy), clarity / understandability (including pronunciation and accent), and audio quality, with any deficiencies addressed.  As the process proceeds, the researcher would continually assess the organization, display, symbol recognition, and accessibility of the content with the participants, addressing deficiencies as necessary. Further, the researcher would determine what, if any, modifications / enhancements to the PictoPages platform may be necessary or desirable in order to provide or improve the intended functionality and whether such changes are feasible within the scope of the project.  Although current mobile devices have few limitations, the researcher needs to be watchful for approaches that may impact response speed, such as excessive dependence on lists of overly complex / high resolution photographic image-based symbols.

\subsection{Validation – with Participants}
Once the content is judged to be essentially complete (subject to corrections and improvements), the researcher will review the use of the aid with the participants and comprehensively test the product with the participants.  Results – which may have both subjective, observational, and measurement-based elements – and participant feedback would be documented.  The results and feedback would be used to correct or improve the application, as required.  This process may require a number of iterations.

\subsection{Validation – ``Real Worl'' (Researcher Observed / Mediated)}
Once both the researcher and the participants are satisfied that the product is working effectively in a ``friendly'' / controlled test environment, the next step would be for the researcher to go into the community with the participants, into various situations, and observe / assist them while they use the product in ``real world'' interactions.  Interactions with individual participants may be pre-planned / structured or free-form or some combination of both.  Although the personal details of such interactions would not be documented, the researcher would need to maintain an awareness to avoid inserting themselves into situations where confidential, personal information may be discussed; the primary area in which this might be of potential concern would probably be with respect to medical situations.  In addition to observing / documenting the participants use of the communication aid, the researcher would also observe and document the response of those with whom the interaction is occurring.  In some cases, those involved may be interested in hearing about the tool and may be willing to provide valuable comments and suggestions (names not sought / recorded).  Based on the results of these initial field trials, the researcher would address any identified problems or deficiencies.

\subsection{Validation – ``Real World'' (Independent) / Final Interviews and Refinements}
As the final validation of the tool, the researcher would acquire, load, and temporarily provide (loan) a number of iPads (or other suitable mobile device) to the participants to use independently for a short time; it may be desirable (and meaningful / important) to extend this opportunity to all those who have participated in the project and contributed to its success.  The researcher would subsequently solicit and document the participants’ feedback, using this to make any final refinements to the product.
 
\subsection{``Marketing'' – Information Dissemination / Encouraging Provision of Access}
In consultation with the partner organization and important stakeholders the Regina Public Library and the Regina Region Local Immigration Partnership,  the product will be publicized and made available to the target community.

\section{Conclusion}
\label{conclusion}

A simple, basic communication aid should assist communication between staff and non-English speaking clients as well as providing assistance to their clients for communicating out in the community. Additionally, having this aid may increase the comfort level of their clients because it gives them a means to tell someone if they have a problem and need help.

To the extent that the communication aid facilitates the efforts of participants to learn English and increases their confidence in communicating with others, it may contribute to their efforts to obtain employment, and reduce their need for financial and other types of assistance from community organizations.  

\section{Acknowledgement}

This research work is supported by MITACS project, Ref. IT12295, and is conducted in collaboration with George Reed Foundation, United Way of Regina and Regina Public Library.

\bibliographystyle{unsrt}

\begin{thebibliography}{8}

 

\bibitem{Bauer}
Visual Media Alliance,  \url {https://main.vma.bz/design/this-new-design-language-is-helping-refugees-in-the-best-way}. Last accessed December, 2020

\bibitem{Pictocom2017}
MAHARAJ, S.:Pictocom International,  \url{http://www.pictoworld.com/presentations.html}. Last accessed December, 2020

\bibitem{Takil2016}
Takil,N-B.: Vocabulary Acquisition with Pictograms and Contextual Sentences in Teaching Turkish as a Foreign Language. Turkish Studies International Periodical for the Languages, Literature and History of Turkish or Turkic \textbf{11/3}(1), 2133--2136 (2016)


\bibitem{clawson2012using}
Clawson, T. H., Leafman, J., Nehrenz Sr, G. M.,Kimmer, S.: Using pictograms for communication. Military medicine  \textbf{177}(3), 291--295 (2012)

\bibitem{Miller2016}
Miller, M.: A Pictogram Language Designed for the Displaced. \url{http://www.fastcompany.com/3063452/a-pictogram-language-designed-for-the-displaced.}. Last accessed November, 2020


\bibitem{McNeill2015}
McNeill, G.: Refugee Communication Board Using Symbols. \url{http://www.callscotland.org/blog/refugee-communication-board-using-symbols.}. Last accessed November, 2020

\bibitem{hartancient}
Hart, A.,  Kindergarten, S. I.: Ancient Forms of Communication in the Spanish Immersion Classroom: The Use of Pictograms and Rebuses to Promote the Development of Spanish Language Literacy.Charlotte Teachers Institute, (2015)

\bibitem{Falck2001}
Falck, K.: The Practical Application of Pictograph. Swedish: Nya Tryckeriet.(2001)
 
\bibitem{SETBC2008} 
SETBC (Special Education Technology – British Columbia).  (2008). Supporting People who use AAC Strategies:  in the Home, School \& Community.  Retrieved from https://www.setbc.org/Download/LearningCentre/Communication/AAC\_Guide\_V4\_Revise\_2008.pdf

\bibitem{Kluzer2013}
Kluzer, S. (2013). Maseltov Deliverable Report D2.1 Immigration and ICT in Europe.  Retrieved from http://www.maseltov.eu/Dokuments/PU\_MASELTOV\_D2.1\_UOC\_2012-07-16\_Immigration-and-ICT-in-Europe\_final.pdf


\end{thebibliography}

\end{document}